\input{aipcheck}

\documentclass[
    ,final            
  ]
  {aipproc}

\newcommand{\be}{\begin{equation}}
\newcommand{\ee}{\end{equation}}
\newcommand{\bea}{\begin{eqnarray}}
\newcommand{\eea}{\end{eqnarray}}
\newcommand{\bem}{\begin{multline}}
\newcommand{\eem}{\end{multline}}
\newcommand{\beg}{\begin{gather}}
\newcommand{\eeg}{\end{gather}}

\newcommand{\as}{\alpha_s}

\def\eq#1{{Eq.~(\ref{#1})}}
\def\fig#1{{Fig.~\ref{#1}}}
\newcommand{\ben}{\begin{eqnarray*}}
\newcommand{\een}{\end{eqnarray*}}

\newcommand{\pd}{\partial}

\graphicspath{{Figs/}}

\layoutstyle{6x9}

\begin{document}

\title{Running Coupling Evolution for Diffractive Dissociation and the NLO Odderon Intercept}


\author{Yuri V. Kovchegov}{
  address={Department of Physics, The Ohio State University, Columbus, OH 43210, USA}
}

\begin{abstract}
  We summarize the results of including running coupling corrections
  into the nonlinear evolution equation for diffractive
  dissociation. We also document a prediction that the NLO QCD odderon
  intercept is zero resulting from a discussion at the Diffraction
  2012 Workshop.
\end{abstract}

\maketitle


\section{Running Coupling Corrections for Diffractive Dissociation}

This proceedings contribution is mainly based on the paper
\cite{Kovchegov:2011aa}.

The evolution equation for single diffractive dissociation in deep
inelastic scattering (DIS) was derived in
\cite{Kovchegov:1999ji}. First we write the cross section for single
diffractive dissociation in DIS on a nucleus as
\begin{equation}\label{xsec}
  M_X^2 \, \frac{d \sigma_{diff}^{\gamma^* A}}{d M_X^2} = - \int d^2
  x_0 \, d^2 x_1 \int\limits_0^1 \, dz \ |\Psi^{\gamma^* \rightarrow q
    {\bar q}} ({x}_{01}, z)|^2 \, \frac{\pd S^D_{{\bf x}_0, {\bf x}_1}
    (Y, Y_0)}{\pd Y_0},
\end{equation}
where $Y= \ln (s/Q^2)$ is the net rapidity interval and the rapidity
gap stretches from rapidity $0$ to rapidity $Y_0 \approx \ln
(s/M_X^2)$ with $M_X^2$ the invariant mass of the produced hadrons.
$|\Psi^{\gamma^* \rightarrow q {\bar q}} ({x}_{01}, z)|^2$ is the
order-$\alpha_{EM}$ light-cone wave function squared for a virtual
photon fluctuating into a $q {\bar q}$ pair with $x_{01} = |{\bf x}_0
- {\bf x}_1|$ the transverse size of the pair and $z$ the fraction of
the longitudinal momentum of the incoming virtual photon carried by
the quark in the pair. The object $S^D$ is the "$S$-matrix" for single
diffractive dissociation, which includes both interacting and
non-interacting contributions in the amplitude and in the complex
conjugate amplitude with the rapidity gap greater than or equal to
$Y_0$ \cite{Kovchegov:2011aa}.

\begin{figure}[ht]
\includegraphics[width=0.8\textwidth]{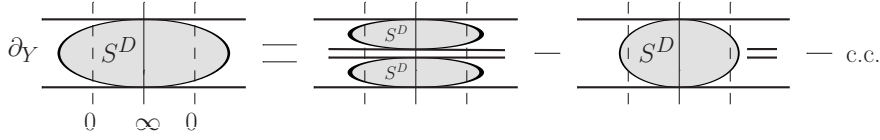}
\caption{The evolution equation for $S^D$.}
\label{SDevol_fig}
\end{figure}

The object $S^D$ obeys a nonlinear evolution equation, which is
equivalent to the Balitsky-Kovchegov (BK)
\cite{Balitsky:1996ub,Kovchegov:1999yj} evolution equation
\cite{Kovchegov:2011aa,Kovchegov:1999ji,Hatta:2006hs}:
\begin{equation}\label{SDevol}
  \pd_Y S^D_{{\bf x}_{0}, {\bf x}_{1}} (Y, Y_0) \, = \, \frac{\as \,
    N_c}{2 \, \pi^2} \, \int \, d^2 x_2 \,
  \frac{x^2_{10}}{x^2_{20}\,x^2_{21}} \, \left[ S^D_{{\bf x}_{0}, {\bf
        x}_{2}} (Y, Y_0) \, S^D_{{\bf x}_{2}, {\bf x}_{1}} (Y, Y_0) -
    S^D_{{\bf x}_{0}, {\bf x}_{1}} (Y, Y_0) \right].
\end{equation}
The equation is illustrated in \fig{SDevol_fig}.
The solid vertical line in \fig{SDevol_fig} denotes the final-state
cut, while the vertical dashed lines denote interactions with the
target nucleus in the amplitude and in the complex-conjugate
amplitude. The initial condition for the evolution \eqref{SDevol} is
given by
\begin{equation}\label{initS}
  S^D_{{\bf x}_0, {\bf x}_1} (Y=Y_0, Y_0) = \left[ 1 - N_{{\bf x}_{0},
      {\bf x}_{1}} (Y_0) \right]^2,
\end{equation}
where $N$ is the (imaginary part) of the forward dipole--nucleus
scattering amplitude obeying the standard BK evolution equation.

It can be shown \cite{Kovchegov:2011aa} that the fact that $S^D$ obeys
the BK evolution results from the cancellation of final state gluon
emissions: no $s$-channel gluon emitted or absorbed after the
interaction with the target (to the right of the dashed line in the
amplitude and to the left of the dashed line in the complex conjugate
amplitude) remains in the final evolution pictured in
\fig{SDevol_fig}.  Without the final state gluon
emissions/absorptions, the evolution becomes just like that for a
forward amplitude, that is, a BK evolution.  The same property remains
valid when the running coupling corrections are included
\cite{Kovchegov:2011aa}: these corrections also cancel in the final
state. Hence, the running-coupling evolution equation for diffractive
dissociation is equivalent to the running-coupling BK (rcBK) evolution
equation, with the initial condition \eqref{initS} now containing the
dipole amplitude $N$ evolved by the full rcBK evolution. We thus write
the running-coupling evolution equation for $S^D$ as
\cite{Kovchegov:2011aa}
\begin{equation}\label{SDevol_rc}
  \pd_Y S^D_{{\bf x}_{0}, {\bf x}_{1}} (Y, Y_0) \, = \, \int \, d^2
  x_2 \, K ({\bf x}_0, {\bf x}_1, {\bf x}_2) \, \left[ S^D_{{\bf
        x}_{0}, {\bf x}_{2}} (Y, Y_0) \, S^D_{{\bf x}_{2}, {\bf
        x}_{1}} (Y, Y_0) - S^D_{{\bf x}_{0}, {\bf x}_{1}} (Y, Y_0)
  \right]
\end{equation}
where the evolution kernel was calculated in
\cite{Kovchegov:2006vj,Balitsky:2006wa}. For completeness let us show
the kernel in the Balitsky prescription \cite{Balitsky:2006wa}:
\begin{equation}\label{kbal}
  K_{rc}^{Bal} ({\bf x}_{0}, {\bf x}_{1}, {\bf x}_{2}) = \frac{N_c \,
    \alpha_s (x_{10}^2)}{2\pi^2} \Bigg[
  \frac{x^2_{10}}{x^2_{20}\,x^2_{21}} +
  \frac{1}{x_{20}^2}\left(\frac{\alpha_s(x_{20}^2)}{\alpha_s(x_{21}^2)}-1\right)+
  \frac{1}{x_{21}^2}\left(\frac{\alpha_s(x_{21}^2)}{\alpha_s(x_{20}^2)}-1\right)
  \Bigg],
\end{equation}
where $\alpha_s (x_\perp^2) = \as \left( 4 \, e^{-\frac{5}{3} - 2 \,
    \gamma_E}/x_\perp^2 \right)$.

\eq{SDevol_rc} can be used to describe the DIS diffraction data with
large center-of-mass energy squared $s$ and large $M_X^2$, such that
$s \gg M_X^2 \gg Q^2$. Unfortunately current HERA data does not extend
to high enough values of $M_X^2$ to necessitate the use of
\eq{SDevol_rc}: perhaps this equation would be useful to describe
single diffraction at the future DIS machines such as the proposed EIC
and LHeC colliders.


\section{The NLO Odderon Intercept}

The progress in the calculation of the next-to-leading order (NLO)
intercept of the QCD odderon was presented in the talk by Jochen
Bartels at the Diffraction 2012 Workshop (reporting on work being done
in collaboration with Victor Fadin and Lev Lipatov). The calculation
employs standard Feynman diagram approach. Here we would like to
document a prediction for the NLO odderon intercept made by the author
of these proceedings in the discussion following the talk: the NLO
odderon intercept can be straightforwardly obtained using the
$s$-channel time-ordered formalism usually employed in saturation
physics.

The odderon exchange amplitude in DIS is \cite{Hatta:2005as}
\begin{equation}
 \label{odd1}
  O_{\bf x \, \bf y} = \frac{1}{2i} \, 
 \frac{1}{N_c}  \left\langle \mathrm{Tr} \, \left[ V_{\bf x}
 \, V^\dagger_{\bf y} \right] -  \mathrm{Tr} \, \left[ V_{\bf y}
 \, V^\dagger_{\bf x} \right] \right\rangle  = \frac{1}{2i} \,  \left[ N_{{\bf y}, {\bf x}} - N_{{\bf x}, {\bf y}} \right]
\end{equation}
where $V_{\bf x}$ is a Wilson line along the light-cone of the
projectile dipole located at transverse coordinate $\bf x$ and
$N_{{\bf x}, {\bf y}} = 1 - \left\langle\mathrm{Tr} \, \left[ V_{\bf
      x} \, V^\dagger_{\bf y} \right]/N_c\right\rangle$ is the
dipole--nucleus forward scattering amplitude.

To construct the NLO evolution equation for $O_{\bf x \, \bf y} $ we
begin with the linearized NLO BK evolution for $N_{{\bf x}, {\bf y}}$
derived in \cite{Balitsky:2008zza} (that is, the NLO
Balitsky-Fadin-Kuraev-Lipatov (BFKL) equation
\cite{Bal-Lip,Kuraev:1977fs} in transverse coordinate space), which
can be written as (see Eq.~(103) in \cite{Balitsky:2008zza})
\begin{equation}\label{Nevol}
\pd_Y N_{{\bf x}, {\bf y}} = \int d^2 z \ K_1 ({\bf x}, {\bf y}; {\bf z}) \, \left[ N_{{\bf x}, {\bf z}} + N_{{\bf z}, {\bf y}} - N_{{\bf x}, {\bf y}} \right] + \int d^2 z \, d^2 z' \, K_2 ({\bf x}, {\bf y}; {\bf z}, {\bf z}') \, N_{{\bf z}, {\bf z}'}.
\end{equation}
The kernels $K_1$ and $K_2$ can be found in Eq.~(103) of
\cite{Balitsky:2008zza}. One can explicitly verify that
\begin{equation}
K_1 ({\bf x}, {\bf y}; {\bf z}) = K_1 ({\bf y}, {\bf x}; {\bf z})  \ \ \mbox{and} \ \  K_2 ({\bf x}, {\bf y}; {\bf z}, {\bf z}') = K_2 ({\bf y}, {\bf x}; {\bf z}', {\bf z}).
\end{equation}
Using these properties of the kernels, along with Eqs.~\eqref{Nevol}
and \eqref{odd1}, we derive the NLO evolution equation for the odderon
amplitude, which turns out to be equivalent to \eq{Nevol}:
\begin{equation}\label{Oevol}
\pd_Y O_{{\bf x}, {\bf y}} = \int d^2 z \ K_1 ({\bf x}, {\bf y}; {\bf z}) \, \left[ O_{{\bf x}, {\bf z}} + O_{{\bf z}, {\bf y}} - O_{{\bf x}, {\bf y}} \right] + \int d^2 z \, d^2 z' \, K_2 ({\bf x}, {\bf y}; {\bf z}, {\bf z}') \, O_{{\bf z}, {\bf z}'}.
\end{equation}
We see that the situation closely replicates that for the odderon
evolution in transverse coordinate space at the leading order (LO)
\cite{Kovchegov:2003dm}: there the odderon evolution equation was also
identical to the (LO) BFKL equation, with the eigenfunctions of the
odderon evolution operator being $C$-odd, that is, they had to flip
sign under the ${\bf x} \leftrightarrow {\bf y}$ interchange. The same
observation applies here at NLO: since the odderon amplitude $O$ obeys
the same NLO BFKL evolution equation, the odderon intercept is the
same as the NLO BFKL intercept, with only odd values of the azimuthal
index $n$ contributing. (Note that, strictly speaking, to obtain
\eq{Nevol} which is equivalent to NLO BFKL one has to redefine the
dipole amplitude by an order-$\as$ correction with the corresponding
new operator referred to as the composite dipole in
\cite{Balitsky:2009xg}: using this operator in place of $N$ would not
change the derivation above.)

The NLO BFKL intercept for non-zero $n$ can be found in
e.g. \cite{Balitsky:2012bs} (see Eq. (65) there).  Evaluating it for
$n=1$ at the saddle point $\gamma = \frac{1}{2} + i \, \nu
=\frac{1}{2}$ we obtain zero\footnote{The fact that the NLO BFKL
  intercept is zero for $n=1$ and $\gamma =1/2$ had been originally
  observed by Agustin Sabio Vera during the discussion of the odderon
  evolution at the Diffraction 2012 Workshop (see also Eq.~(38) in
  \cite{Vera:2006un}), and was since confirmed by the author.}, such
that the NLO odderon intercept in QCD is
\begin{equation}\label{Oint}
\alpha_O - 1 = 0 + O(\alpha^3_s). 
\end{equation}
We conclude that the zero value of the odderon intercept, originally
found at the leading-order in \cite{Bartels:1999yt}, persists at the
NLO.\footnote{Note that Mikhail Braun showed that running-coupling
  corrections do not change the LO odderon intercept, leaving it at
  zero \cite{Braun:2007kz}: our result \eqref{Oint} is consistent with
  this conclusion.}

It is interesting to note that, as was pointed out by Lipatov in the
discussion at the Workshop, the odderon intercept at strong 't Hooft
coupling $\lambda = g^2 \, N_c$ was found in \cite{Brower:2008cy}
using the Anti-de Sitter space/Conformal Field Theory (AdS/CFT)
correspondence. The authors of \cite{Brower:2008cy} found two odderon
solutions, both of which give $\alpha_O -1 \rightarrow 0$ for $\lambda
\rightarrow \infty$, with one of the solutions exhibiting no deviation
from zero when finite-$\lambda$ corrections were calculated, such that
$\alpha_O -1 =0 + O(1/\lambda)$. This result, along with our
\eq{Oint}, presents evidence in favor of a tantalizing possibility
that the odderon intercept is identically equal to zero at all values
of the coupling! (The same reference \cite{Brower:2008cy} states that
the result \eqref{Oint} had been known earlier to Cyrille Marquet,
who, as it turns out, arrived at it numerically using a similar line
of arguments to the one presented here.)
  

\vspace*{-1mm}

\begin{theacknowledgments}
  The author is greatly indebted to Ian Balitsky and Giovanni Chirilli
  for discussions of the NLO BK evolution. This work is sponsored in
  part by the U.S. Department of Energy under Grant No. DE-SC0004286.
\end{theacknowledgments}

\vspace*{-1mm}





\end{document}